# Predictive Analysis for Social Diffusion: The Role of Network Communities

Richard Colbaugh[*] and Kristin Glass[†]

[*] rcolbau@sandia.gov, Sandia National Laboratories, Albuquerque, NM USA
[†] kglass@icasa.nmt.edu, New Mexico Tech, Socorro, NM USA

**Abstract.** The diffusion of information and behaviors over social networks is of considerable interest in research fields ranging from sociology to computer science and application domains such as marketing, finance, human health, and national security. Of particular interest is the possibility to develop predictive capabilities for social diffusion, for instance enabling early identification of diffusion processes which are likely to become "viral" and propagate to a significant fraction of the population. Recently we have shown, using theoretical analysis, that the dynamics of social diffusion may depend crucially upon the interactions of *social network communities*, that is, densely connected groupings of individuals which have only relatively few links between groups. This paper presents an empirical investigation of two related hypotheses which follow from this finding: 1.) inter-community interaction is predictive of the reach of social diffusion and 2.) dispersion of the diffusion phenomenon across network communities is a useful early indicator that the propagation will be "successful". We explore these hypotheses with case studies involving the emergence of the Swedish Social Democratic Party at the turn of the 20th century, the spread of the SARS virus in 2002-2003, and blogging dynamics associated with real world protest activity. These empirical studies demonstrate that network community-based diffusion metrics do indeed possess predictive power, and in fact can be more useful than standard measures.

## 1 Introduction

The diffusion of ideas, opinions, innovations, and behaviors over social networks is of considerable scientific and practical importance. Of particular interest is the possibility to develop *predictive* capabilities for social diffusion. Enormous resources are devoted to the task of predicting the outcomes of diffusion processes, in domains ranging from public policy to popular culture to national security, but the quality of such predictions is often quite poor. It is tempting to conclude that the problem is one of insufficient information. Clearly diffusion phenomena which "go viral" are qualitatively different from those which quickly fade into obscurity or they wouldn't be so dominant, the conventional wisdom goes, so in order to make good predictions we should collect enough data to allow these crucial differences to be identified.

Recent research in the social and behavioral sciences calls into question this conventional wisdom and, indeed, indicates that there may be fundamental limits to what can be predicted about social diffusion. For example, the studies reported in [1,2] indicate that the intrinsic attributes normally considered to be important when assessing the likelihood of diffusion success, such as the appeal of the product in a word-of-mouth advertising campaign, do not possess much predictive power. Such research provides evidence that for social diffusion processes, in which individuals are influenced by what others do, it is not possible to obtain reliable predictions using standard methods which focus on the intrinsic characteristics of potential process outcomes. We propose that accurate prediction, if it is possible at all, requires careful consideration of the subtle interplay between the intrinsics of the process and the underlying diffusion dynamics which is its realization.

As a first step toward this goal we have recently conducted a theoretical analysis of social network diffusion [3]. The analytic framework employed in this study is inspired by recent work in biology demonstrating that the stochastic hybrid dynamical system (S-HDS) is a useful mathematical formalism with which to represent multiscale biological network dynamics [4]. Using this theoretical framework we have shown that the dynamics of social diffusion may depend crucially upon the interactions of *social network communities*, that is, densely connected groupings of individuals which have only relatively few links between groups [5]; see Figure 1 for a illustration of the basic concept. Recall that community structure is ubiquitous in real world social networks [5].

This paper presents an empirical examination of two hypotheses which follow from the theoretical work reported in [3]: 1.) inter-community interaction is predictive of the reach of social diffusion and 2.) dispersion of the diffusion phenomenon across network communities is a useful early indicator that the propagation will be "successful". We explore the first hypothesis with case studies of the emergence of the Swedish Social Democratic Party at the turn of the 20th century and the spread of SARS in 2002-2003. The second hypothesis is investigated through analysis of blogging dynamics associated with protest activity. These empirical studies demonstrate that network community-based diffusion metrics do possess predictive power, and indeed can be more useful than standard measures.

## 2 Social Diffusion and Network Communities

In social diffusion processes, individuals are affected by what others do. This is easy to visualize in the case of disease transmission, with infections being passed from person to person. Behaviors and innovations can also propagate through a population, as individuals imitate others in an attempt to obtain the benefits of coordinated action, infer otherwise inaccessible information, or manage complexity in decision-making. The dynamics of social diffusion depends upon the topological features of the underlying social network, for instance the degree distribution or presence of small world structure, and aspects of this dependence have been characterized (see [6] for a recent review). Interestingly, although network community structure is ubiquitous in real world social network, little has been done to quantify the role of communities in diffusion phenomena. One reason for this gap in understanding could be that standard network analysis methods, such as those borrowed from statistical physics and econometrics, are not well-suited for investigating networks with community structure.

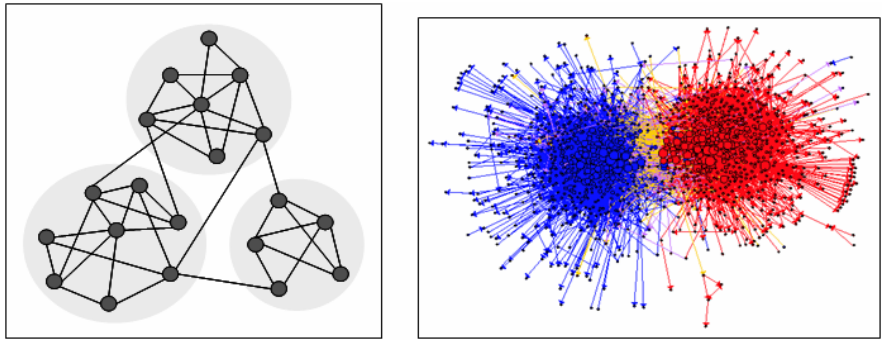

**Fig. 1.** Network community structure. Cartoon at left depicts a network with three communities; graph at right is a network of political blogs [7].

In [3] we a present a theoretical analysis of social diffusion on networks with realistic topologies, including community structure. The analysis leverages S-HDS models for network dynamics. An S-HDS is a feedback interconnection of a discrete-state stochastic process, such as a Markov chain, with a family of continuous-state stochastic dynamical systems. Combining discrete and continuous dynamics within a single computationally tractable framework provides an expressive, scalable modeling environment that is both intuitively accessible and amenable to formal analysis. In particular, the analytic framework developed in [3] enables the study of social dynamics on networks of real world scale and complexity. A focus of [3] is the derivation of provably-correct characterizations of diffusion processes on networks with realistic degree distributions, small world structure, community structure, and exogenous inputs (e.g., from traditional media).

Using this theoretical framework in combination with empirically-grounded models for social diffusion [e.g., 6,8], we have demonstrated that the dynamics of these diffusion models depends crucially upon the social network communities. More specifi-cally, one of the analytic conclusions of this study can be expressed in terms of two hypotheses.

**Hypothesis 1:** The "intensity" of inter-community interaction is an important predictor of the magnitude and reach of social diffusion.

**Hypothesis 2:** The "dispersion" of a diffusion phenomenon across disparate network communities is a useful early indicator that propagation will be significant.

This paper presents the results of an empirical assessment of (more precise versions of) these two hypotheses. We explore Hypothesis 1 through two case studies, the first involving the emergence and diffusion of the Swedish Social Democratic Party at the turn of the 20th century and the second focusing on the spread of SARS during 2002-2003. Hypothesis 2 is investigated through an additional case study, involving mobilization/protest activity and the associated online diffusion of protest-relevant information. These case studies provide considerable support for Hypotheses 1 and 2, for example showing that network community-based diffusion metrics do indeed possess predictive power, and in fact can be more useful than standard measures.

## 3 Empirical Assessment of Hypothesis 1

We begin our assessment of Hypothesis 1 with a case study involving the Swedish Social Democratic Party (SDP). The SDP was founded in Stockholm in 1889 and grew to become one of the most successful political parties in the world. During the early years of the SDP, party activists traveled throughout Sweden attempting to generate interest and support. One consequence of this activity was the formation of a network of "long range connections" linking previously disparate communities [8]. Thus early diffusion of the SDP took place over a social network composed of geographically and demographically-based network communities connected to each other by a sparse set of inter-community links (see Figure 2). Because both the membership dynamics and activist activities are well-documented [8,9], SDP diffusion over this "network of networks" provides a convenient basis for an empirical assessment of Hypothesis 1.

Consider the evolution of SDP membership in the 369 jurisdictional districts of Sweden during 1889-1918 (the critical early years of party growth [8]). Figure 2 illustrates the basic features of this diffusion process. The visualization at the right of the figure depicts the temporal evolution of concentration of SDP members for all districts, with the vertical axis indicating district index, the horizontal axis corresponding to time, and the colors indicating variation from minimum member concentration (dark blue) to maximum concentration (red); the map on the left of the figure shows a few representative inter-community links. Time series analysis in [8] shows that intrinsic characteristics of SDP diffusion, such as the appeal of certain activists and

Predictive Analysis for Social Diffusion

demographics of the local regions, are not predictive of membership growth; however, this analysis also suggests that membership levels in the K districts which form the vertices of the activist-induced inter-community network may possess predictive power. Thus we turn to a social dynamics-oriented analysis, and examine the predictive power of the following simple model for party diffusion:

$$\Sigma_{SDP}: \quad m_i(t+1) = \alpha + \beta_1 m_i(t) + \beta_2 m_m(t) + \beta_3 m_{i,\ geo\text{-}m}(t) + \beta_4 m_{net\text{-}m}(t),$$

where t is time in years, $m_i(t)$ is party membership concentration (MC) in district i at time t, $m_m(t)$ is (country-wide) mean MC, $m_{i,\ geo\text{-}m}(t)$ is mean MC for the geographic region in which i is located [9], $m_{net\text{-}m}(t)$ is mean MC for the K districts which define the inter-community network, and $\{\alpha, \beta_1, \beta_2, \beta_3, \beta_4\}$ are the model parameters.

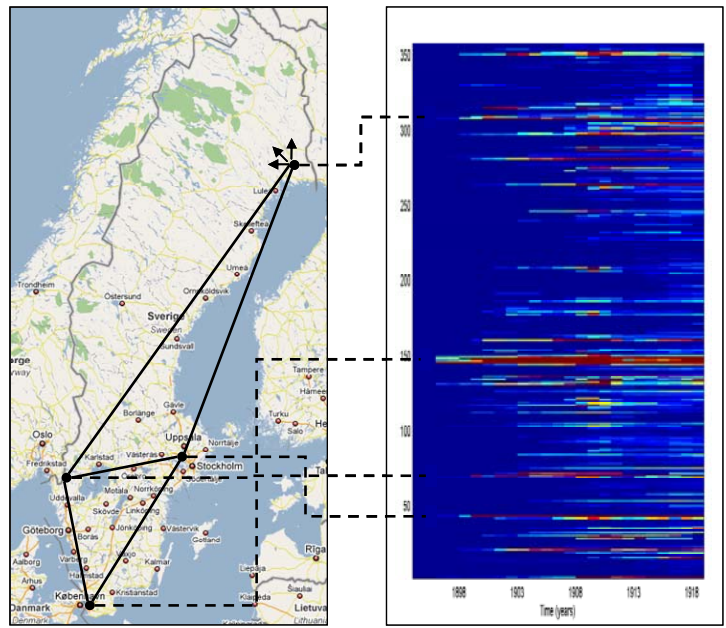

**Fig. 2.** Illustration of SDP inter-community interaction network. Map at left shows a few representative links between previously disparate, geographically-based network communities. Visualization at right depicts temporal evolution of concentration of SDP membership in 369 jurisdictional districts during the period 1889-1918.

Estimating the model $\Sigma_{SDP}$ using data obtained from [8,9] for the K districts which possess both intra-community and inter-community interactions reveals that: 1.) all of the coefficients $\{\beta_1, \beta_2, \beta_3, \beta_4\}$ are predictive of party membership growth ($p<0.05$) and 2.) $\beta_4 \, m_{net-m}$ is the most predictive term in the model. Observe that the latter result is somewhat surprising: inter-community interaction is more predictive of SDP membership for a given district for the next year than the current membership level for that district, and is much more predictive than membership levels in geographically adjacent districts. Thus inter-community interaction is more predictive than either process intrinsics or standard diffusion measures (which focus on intra-community dynamics).

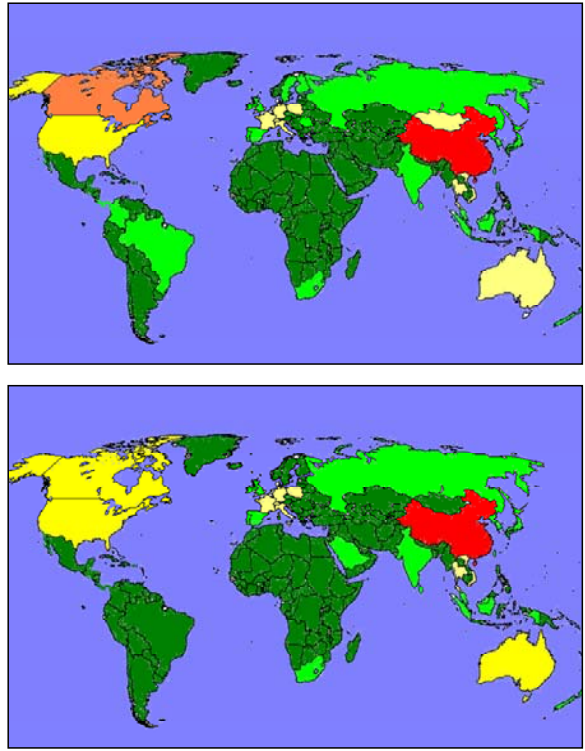

**Fig. 3.** Geographic spread of SARS epidemic: actual (top) and simulated (bottom) cumulative SARS infection levels by country (dark green is lowest, red is highest).

The next case study illustrates the way Hypothesis 1 can be used to guide development of simple but still useful models for diffusion phenomena. Consider the outbreak and rapid propagation of severe acute respiratory syndrome (SARS) in 2002-2003. To obtain a very simple model for the global spread of SARS, we define each country to be a social network community and represent intra-community disease transmission with a stochastic susceptible-infected-removed model for a "fully mixed" population [10]. Inter-community interactions are assumed to consist of individuals traveling between countries along commercial air travel routes (data on air travel was obtained from [11]). This characterization of inter-community dynamics results in a model which is easily and rapidly constructed from publicly available data. Moreover, despite its simplicity, the model provides a useful description for the spread of SARS. For example, Figure 3 shows that simulations of the model are in good agreement with the actual spread of the SARS epidemic during 2002-2003. Observe that the utility of this simple model is a consequence of allocating modeling detail according to Hypothesis 1: because inter-community interaction is a key element of social network diffusion, this dynamics is modeled more carefully than intra-community interaction. Additional details regarding the SARS model summarized here are given in [12].

## 4 Empirical Assessment of Hypothesis 2

This case study examines whether "dispersion" of a diffusion process across disparate social network communities is a useful early indicator for successful mobilization and protest events. The investigation focuses on Muslim reaction to nine recent incidents, each of which appeared at the outset to have the potential to trigger significant protests. This collection of events includes publication of cartoons depicting Mohammad in the Danish newspaper *Jyllands-Posten* in September 2005, the lecture given by Pope Benedict XVI in September 2006 quoting controversial material concerning Islam, and republication of the "Danish cartoons" in various newspapers in February 2008. Recall that the first Danish cartoons event ultimately led to substantial Muslim mobilization, including massive protests and considerable violence, while in contrast Muslim outrage triggered by the pope lecture and the second Danish cartoons event subsided quickly with essentially no violence. The set of nine events contains four which became large and self-sustaining and five that quickly dissipated, so that taken together these events provide a useful setting for testing whether the extent of early diffusion across social communities can be used to distinguish nascent mobilization events which will become successful from those that will not.

A central element in the proposed approach to early warning analysis is the measurement, and appropriate processing, of social dynamics associated with the process of interest. In the present study we use *online* social activity as a proxy for real world diffusion of mobilization-related discussions and information. More specifically, we use blog posts as our primary data set. The "blogosphere" is modeled as a graph composed of two types of vertices, the blogs themselves and the concepts which appear in them. Two blogs are linked if a post in one hyperlinks to a post in the other, and a blog is linked to a concept if the blog contains (significant) occurrences of that concept. Among other things, this blog graph model enables the identification of blog commu-

nities – that is, groups of blogs with intra-group edge densities that are significantly higher than expected; these blog communities serve as a proxy for social communities.

We propose the following procedure for protest warning analysis using blog data:

Given a potential triggering event of interest

1. Perform a focused blog graph crawl, with a lexicon constructed for the triggering event of interest, to collect relevant blog posts and build the associated blog graph.
2. Partition the blog graph into network communities.
3. Assemble post volume time series and compute post/community entropy (PCE) time series associated with the post volume dynamics.
4. Construct a synthetic ensemble of PCE time series from the post volume dynamics.
5. Motif detection: compare actual PCE time series to the synthetic ensemble series to determine if the early diffusion of activity across communities is "excessive".

We now provide a few additional details concerning this procedure. Step 1 is by now standard, and various off-the-shelf tools exist which can perform this task [12]. In Step 2, blog graph communities are identified through a standard community extraction algorithm applied to the blog graph [5,12,13]. In Step 3, post volume for a given community i and sampling interval t is obtained by counting the number of relevant posts made in the blogs comprising community i during interval t. PCE for a given sampling interval t is defined as follows:

$$PCE(t) = -\Sigma_i f_i(t) \log(f_i(t)),$$

where $f_i(t)$ is the fraction of total relevant posts made during interval t which occur in community i. Given the post volume time series obtained in Step 3, Step 4 involves the construction of an ensemble of PCE time series that would be expected under "normal circumstances", that is, if Muslim reaction to the triggering event diffused from a small "seed set" of initiators according to the S-HDS social dynamics model introduced in [3]. Finally, Step 5 is carried out by searching for time periods, if any, during which the actual PCE time series exceeds the mean of the synthetic PCE ensemble by more than two standard errors.

Illustrative time series plots associated with the proposed approach to mobilization/protest early warning are shown in Figure 4. Observe that in the case of the first Danish cartoons event the PCE of relevant discussions (blue curve) experiences a dramatic increase a few weeks before the corresponding increase in volume of blog discussions (red curve); this latter increase, in turn, takes place before any violence. In contrast, in the case of the pope event, PCE of blog discussions is small relative to the cartoons event, and any increase in this measure lags discussion volume. Similar time series plots are obtained for the other seven events, suggesting that early diffusion of discussions across blog communities may be a useful indicator of large mobilization events.

Predictive Analysis for Social Diffusion

To examine this possibility more carefully, we apply three early warning methods to our set of protest events: 1.) the motif detection algorithm summarized above, 2.) a simple volume-based technique, in which significant post volume is used to predict which events will become large and which will dissipate, and 3.) a random early warning method, in which predictions of large/small protests are assigned at random (with the correct probability distribution for large/small events). We find that the motif detection method provides statistically significant early warning accuracy ($p<0.005$) while the volume-based method does not (it is not significant at even the $p=0.1$ level). This case study therefore provides empirical support for Hypothesis 2, indicating that early diffusion of mobilization-related activity (here blog discussions) across disparate social communities can be a useful early signature of successful protest events.

.

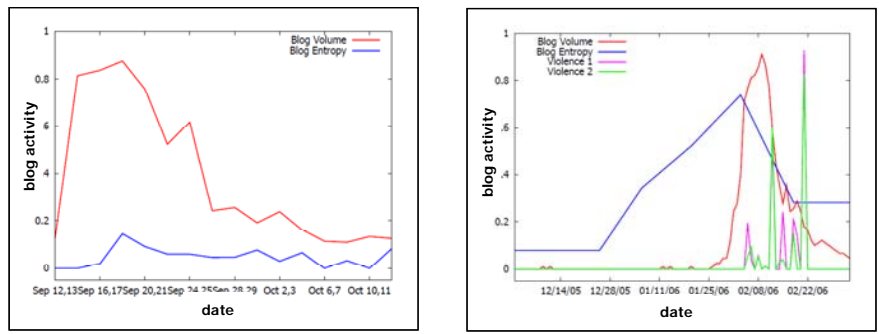

**Fig. 4.** Sample results for mobilization/protest case study. The illustrative time series plots shown correspond to the pope event (left) and first Danish cartoons event (right). In each plot, the red curve is blog volume and the blue curve is blog entropy; the Danish cartoon plot also shows two measures of violence (cyan and magenta curves). Note that while the volume and violence data are scaled to allow multiple data sets to be graphed on each plot, the scale for entropy is consistent across plots to enable cross-event comparison.

**Acknowledgements.** This research was supported by the U.S. Department of Homeland Security, the U.S. Department of Defense, and the Laboratory Directed Research and Development program at Sandia National Laboratories. Discussions on aspects of this work with Paul Ormerod of Volterra Consulting are gratefully acknowledged.